# Boosting transducer matrix sensitivity for 3D large field ultrasound localization microscopy using a multi-lens diffracting layer: a simulation study

Hugues Favre[1], Mathieu Pernot[1] Mickael Tanter[1] and Clément Papadacci[1]

[1] Institute Physics for Medicine Paris, Inserm U1273, ESPCI Paris-PSL, Cnrs UMR8063, 75012 Paris, France

E-mail: clement.papadacci@espci.fr

## Abstract

Mapping blood microflows of the whole brain is crucial for early diagnosis of cerebral diseases. Ultrasound localization microscopy (ULM) was recently applied to map and quantify blood microflows in 2D in the brain of adult patients down to the micron scale. Whole brain 3D clinical ULM remains challenging due to the transcranial energy loss which significantly reduces the imaging sensitivity. Large aperture probes with a large surface can increase both resolution and sensitivity. However, a large active surface implies thousands of acoustic elements, with limited clinical translation. In this study, we investigate via simulations a new high-sensitive 3D imaging approach based on large diverging elements, combined with an adapted beamforming with corrected delay laws, to increase sensitivity. First, pressure fields from single elements with different sizes and shapes were simulated. High directivity was measured for curved element while maintaining high transmit pressure. Matrix arrays of 256 elements with a dimension of 10x10cm with small (λ/2), large (4λ), and curved elements (4λ) were compared through point spread functions analysis. A large synthetic microvessel phantom filled with 100 microbubbles per frame was imaged using the matrix arrays in a transcranial configuration. 93% of the bubbles were detected with the proposed approach demonstrating that the multi-lens diffracting layer has a strong potential to enable 3D ULM over a large field of view through the bones.

## 1. Introduction

Blood circulation is essential to the organs' functions and occurs through a complex network of vessels with diameters varying from several millimeters for large arteries down to only a few microns for capillaries. In many diseases, microcirculation architecture and its functional abnormalities are key markers of the outbreak of the disease. Many pathologies however remain diagnosed only at later stages, when observable symptoms become visible at larger scales. Assessing the blood flows across several spatial scales (down to the micron scale across the whole organ) is therefore essential for early diagnosis but remains a major challenge in clinical medical imaging.

MRI and CT scanners offer effective whole-organ or whole-body imaging, but mostly provide an anatomical mapping with a spatial resolution limited to a few millimeters. Besides, their cost and cumbersomeness restrict their accessibility for patients. Optical intravital microscopy is considered the gold standard for in vivo microcirculation assessment of thin tissue which allows transillumination and human studies have been limited to observation of nail fold capillaries. In general, optical-based imaging techniques mostly suffer from low tissue penetration which limits the clinical applications.

Ultrasound is increasingly used to assess blood flows in clinics but usually suffers from low resolution and sensitivity, especially when imaging through bones. Ultrasound imaging is however an extremely active research field with many technical advances achieved through the past decades, which can rapidly be translated to clinics given the relatively low cost and portability of ultrasound technologies.

In the last decade, the principle of optical localization microscopy (stochastic optical reconstruction microscopy (STORM), photoactivated localization microscopy (PALM), and fluorescence PALM (fPALM)) was translated to the ultrasound domain, resulting in a new non-invasive method for assessing microflows at depth in organs: Ultrasound Localization Microscopy (ULM) maps blood vessels and their dynamics at resolutions as small as ten micrometers [1]. ULM breaks the fundamental ultrasound diffraction limit over an order of magnitude at depths much greater than the traditionally frequency-limited imaging depth [2]. ULM combines ultrafast ultrasound imaging with contrast agents injected into the bloodstream. These contrast agents are commercially available microbubbles for sonography and are routinely used in clinical practice [3]. Leveraging ultrafast imaging rates (typically 10 000 frames/sec), the circulating contrast agents are tracked individually and their centers are localized beyond the diffraction-limited spatial resolution to reconstruct vascular maps with a high spatial resolution. ULM was first applied to reveal cerebral, tumoral, and kidney microflows on animal models [1,4–6]. In the rat brain, the spatial resolution was increased by a factor of 50 compared with conventional Doppler imaging. Clinical feasibility of cerebral microflow imaging was demonstrated in a study published in 2021 [7]: ULM was applied to image patients with a cerebral aneurysm, revealing the human cerebral vascular anatomy and microflow dynamics for the first time at microscopic scales in 2D.

However, 2D ultrasound imaging increases the risk of missing out-of-planes features during diagnostic examinations in pathologies associated with complex spatial architecture such as glioblastoma multiforme or arteriovenous malformation. It can also be difficult to perform longitudinal follow-up on cross-sections acquired in different imaging sessions. Finally, 2D imaging is highly operator-dependent because probe positioning is critical to capture the appropriate cross-sectional view. Imaging the whole organ in three dimensions is an important aspect in clinical practice for abnormalities detection, as well as in fundamental research to fully characterize the organ's function. Preliminary developments of 3D ULM have recently been initiated with the development of the first programmable 3D ultrafast ultrasound scanners capable of acquiring up to 5000 volumes/sec [8–15]. The feasibility of 3D ultrafast ULM was demonstrated in microtubes [16–21].

Capturing large 3D volumes at a high volume rate through bones such as the skull remains challenging in ULM, and more generally in the field of medical ultrasound imaging. Moreover, the acoustic energy loss due to the reflection and attenuation of ultrasound waves at the bone interface significantly reduces the image sensitivity. Strategies involving small aperture probes were developed to image the brain or the heart through favorable acoustic windows such as the temporal bone (where the skull is thinner) [22] or the intercostal space (no bone). However, small apertures limit both the field of view and the spatial resolution. On the other hand, large aperture matrix probes with thousands of acoustic elements imply heavy technological developments and tremendous cost with limited clinical translation. For all these reasons, researchers in the field proposed different approaches to decrease the number of elements while maintaining a large aperture. Sparse array techniques [23] consist in under-sampling the aperture by sparsely distributing a limited number of small elements across the probe surface. This new concept allowed performing volumetric imaging with promising performances [21,24]. However, small elements (<1x1$\lambda$) limit the acoustic energy and are therefore not be suitable for imaging through bones.

To maintain a large active surface as well as reduce the number of elements, row-column addressed (RCA) arrays have been developed recently [25–28]. The field of view is classically limited to the surface of the probe in the lateral dimension but can be increased using a convex shape or a diverging lens [29,30]. However, ultrafast imaging requires a small number of emissions which induces high secondary lobes and limited sensitivity when applied to RCA arrays [31]. Moreover, for transcranial application, a large probe surface (>5x5cm²) is required, therefore the dimension of the transducers in one dimension would be necessarily much larger than the coherence length of the skull (1cm at 1MHz) [32,33], which would induce aberration blurring and challenging aberration correction methodologies.

Another intuitive approach consists in using sparse large elements to increase emitted and received energy signals, but this approach suffers from a low angular directivity, degrading the field of view and the image quality due to high grating lobes.

Negative acoustic lenses were introduced in the field of photoacoustic imaging, where there is a need for detecting extremely weak acoustic signals while preserving a large field of view [34]. Another group recently developed a multi-lens array and performed receive beamforming to successfully increase acoustic signal in 3D photoacoustic applications [35].

In this study, we propose a new approach based on the concept of multi-lens probes with a reduced number of large elements driven at a low acoustic frequency to perform ultrasensitive 3D ULM of the entire brain through the skull bone (figure 1). The approach will be implemented and studied through simulations. Firstly, a negative acoustic lens will be modeled numerically by a curved transducer and the effect of transducer size on the acoustic field will be studied and compared to flat transducers. Secondly, acoustic fields from three matrix array probes: sparse arrays with small, large, and curved elements will be compared. Thirdly, ultrafast imaging will be applied to a static phantom with microbubbles to determine the imaging capabilities of the three different configurations in a large 3D field of view. Finally, a realistic vasculature phantom with flowing bubbles will be imaged to demonstrate the feasibility of performing 3D ultrafast ULM through the skull in a large field of view with a multi-lens matrix probe.

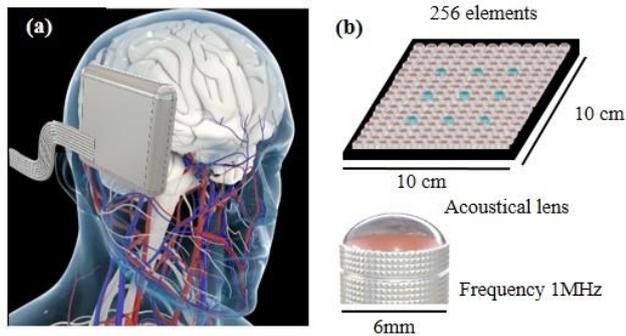

Figure. 1. Ultrasensitive probe with 256 large diverging elements for 3D large field transcranial ultrasound localization microscopy of the brain. (a) Illustration of the probe application for cerebral microcirculation imaging. (b) Illustration of the probe with the surface of 10x10cm², composed 256 circular elements of 6mm diameter, with multi-lenses. Blue transducers are used in transmit mode.

## 2. Materials and methods

*2.1 Simulation models*

*2.1.1 Criteria for transcranial microbubble imaging and simulation approach.*

Single element behavior, array effects, static phantom imaging, and realistic dynamic flow phantom were studied in simulations. The ultrasound simulations were performed in Matlab using Field II [36,37]. Comparisons in terms of pressure amplitude, directivity, and image quality were performed. To compare the different configurations, we assumed the pressure amplitude at the surface of the transducers to be the same for all the configurations in transmit mode. This assumption relies on the fact that for equivalent pulse durations, safety limitations are mainly linked to the pressure amplitude emission which includes, MI, ISPTA, and bone heating consideration [38]. The

TABLE 1

| Parameter | Value |
|---|---|
| Frequency | 1 (MHz) |
| Bandwidth | 66% |
| Number of cycle | 2 |
| Sound of speed (c) | 1480 (m.s$^{-1}$) |
| Wavelength (λ) | 1.5 (mm) |
| λ/2 element size | 0.75 (mm) |
| Flat 4λ element size | 6 (mm) |
| Curved 4λ element size | 6 (mm) |
| Curvature of curved elements (α) | 40° |
| Apodization | Tukey, r=1/3 |
| Spatial frequency | 0.75 (mm) |
| Sampling frequency | 20 (MHz) |

Table1: Simulation parameters

frequency of the transducers was chosen to be 1MHz to limit the attenuation coefficient of the wave in the skull bone (10dB/cm), which scales in $f^{2.1}$ around 1MHz [39]. It was also a choice of not going below 1MHz for resolution purposes (table 1). At 1MHz frequency, the wavelength corresponds to 1.5mm considering a speed of sound of 1480m/s. The coherent length of the human skull is typically around 1cm at 1MHz [32,33], therefore, to avoid any spatial undersampling of the skull aberration law, we consider a maximum transducer length of 6mm (4λ) in the simulation study.

*2.1.2 Transducer modeling.*

To summarize, three types of cylindrical 3D transducers were compared. 1) Transducer with a diameter of λ/2 with low directivity, where λ is the wavelength. 2) Transducer with a diameter of 4λ to maximize energy in transmit and receive mode. 3) Diverging transducer with a diameter of 4λ modeled by a curved transducer. The element curvature was defined as the angle between the base of the curved transducer and the tangent at the edge (figure 2). An apodization on the transducer's surface was added both in emission and in reception mode. The apodization function is a Tukey window with a cosine fraction r of 1/3. The transducers were simulated as common resonant piezoelectric ceramics with a bandwidth of 66%. Parameters used in the simulations are displayed in table 1.

*2.2 Characterization of a single element*

The objective of this part is to determine the performances of the three types of elements: small, large, and large diverging elements. Thus, pressure fields of flat and curved single element transducers were simulated at a diameter size increment of λ/2 up to 5λ. Pressure fields were computed as the maximum amplitude over time of the transmit impulse response at each point of space of a volume with dimension 10cm x 10cm x 10cm centered on the single element in the axial plane. Pressure fields were normalized by the maximum in the field of view and a log compression was applied to display volumes.

From pressure field simulations, Gain quantifications were calculated as a function of transducer diameter for flat and curved transducers. Amplitude Gains (G) were estimated at two points of the volume using the formula (1), where d is the transducer diameter. The first point was defined by the spherical coordinates (r=10cm, θ=0°, φ=0°) which correspond to a point in front of the transducer. A second point was defined off-axis by the coordinates (r=10cm, θ=30°, φ=0°). Directivity was quantified along a radial amplitude profile of radius 10 cm, as the angle defined by a -20dB pressure drop.

$$G_{r,\theta}(d) = \frac{\max[Amplitude_{r,\theta}(d)]}{\max[Amplitude_{r,\theta}(d=\lambda/2)]} \quad (1)$$

*2.3 Multi-element array comparison*

*2.3.1 Array definition.*

In this study, three matrix arrays with surfaces of 10x10cm² were filled with 256 elements distributed with a 4λ pitch in x and y directions. The first and the second array were made of λ/2 and 4λ diameter elements, respectively. The third array was composed of 4λ diameter curved elements. In addition, a reference array with a surface of 10x10cm² was filled with 16900 elements of diameter λ/2 distributed with a λ/2 pitch in x and y directions.

*2.3.2 Focusing delays.*

For the first and second arrays, a classical focusing delay law was applied by computing the time delay differences corresponding to the Euclidian distance between the center of the transducer C and the focal point P (Eq2).

$$\tau(P) = \frac{dist(C,P)}{c} \quad (2)$$

For the curved elements, however, an additional delay must be taken into account to correct the curvature effect. The delay correction for the matrix composed of curved transducers was applied according to formula (3). The distance is computed from the focal point F of the element to the point of interest P. For a curved element, F is the center of the circle arch defined by the curvature of the transducer (figure 2(a)). For a flat transducer with a lens, F is the focal point of the lens, which depends on the speed of sound in the lens material and the lens curvature. The distance between the Focal point F and the intersection point of the segment [FP] and the curved interface is called r. The distance r is subtracted to the previous distance [F,P] and then divided by the velocity of the medium c to compute the 'corrected' delay (equation 3).

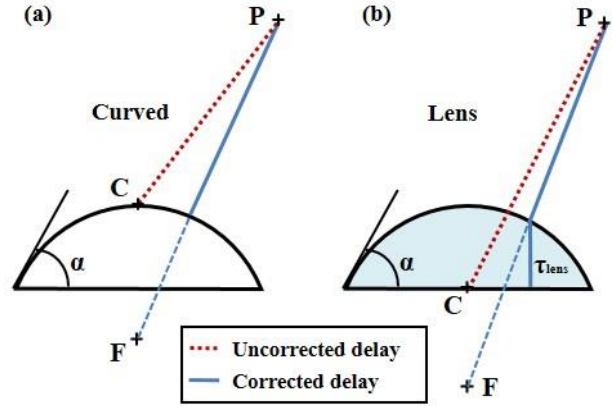

Figure. 2. Schematic of the curved transducer, and the delay traces. Alpha defines the curvature of the transducer, it is the angle between the horizontal and the tangent at the edge of the transducer. P is the point of interest in the imaging field, C is the center of the transducer, and F is the focal point. (a) Case of a large curved element. (b) Case of a flat element with a diverging acoustic lens.

$$\tau(P) = \frac{dist(F,P) - r}{c} + \tau_{lens} \quad (3)$$

Note that for a diverging acoustical lens, the delay $\tau_{lens}$ must be added as the path inside the lens needs to be taken into account (figure 2(b)). For curved elements, $\tau_{lens}$ is set to 0.

Maximal pressure field from focus transmit was assessed for the three different arrays at the focal point located at [0, 0, 5cm]. The imaging field dimension was set to 10 by 10 by 10 cm. Pressure fields were normalized by the maximum in the field of view and a log compression was applied to display volumes. The profiles at the focus depth were normalized by the maximum pressure at the focus obtained with the array composed of λ/2 elements.

In addition, the pressure field profile obtained with the reference array was also assessed.

*2.3.1 Ultrafast imaging.*

In this study, to perform ultrafast imaging with coherent compounding, N=9 transmits per frame are used. The 9 transmits are performed by 9 different elements of the probe emitting successively. The first emitting element is located at the center of the array. The others are located at a distance of +/- 24mm around the center, as shown in figure 1(b). In the receive mode, dynamic focusing was performed in each voxel with a constant F/D value of 1 (where F is the pixel depth and D is the aperture size) using all the elements.

After each transmit, the simulated backscattered signals (RF) were acquired by the 256 elements of the probe. The signals of each element were processed using a 3D delay and sum algorithm with delay correction for the curved elements. Coherent compounding was performed by coherently summing the 9 images.

*2.3.1 Point spread function.*

3D Point Spread Function (PSF) for a point scatter located at [0, 0, 5cm], in cartesian coordinate, was compared for the three matrix arrays. Ultrafast imaging with coherent compounding, dynamic focusing, and beamforming was used as described above. 3D PSF were normalized by their maximum and log compression was applied. The projection of the maximum value on the Oxy plane was displayed as well as the orthogonal projection of the maximum values on the x-axis.

*2.4 Static and particle flow phantoms*

*2.4.1 Static phantom.*

Microbubbles were modeled as randomly distributed point scatterers in the simulations, with a reflective coefficient set to one. A static phantom was created to assess the image quality of a frame using the three different matrices. N=200 bubble positions were randomly chosen in a volume with dimensions of 4 by 4 by 9 cm, centered on the probe. The images obtained by the three matrix arrays were compared and displayed in log compression. The plane Oxz and the x profile of a bubble located at [0, 0, 5.4cm] was displayed for the three cases.

To model transcranial energy loss, RF amplitude signals had been attenuated by 20dB, corresponding to twice the passage of 1cm of skull bone [39]. The transmitted pressure was set to 1MPa and a random noise with a 100Pa equivalent noise pressure was added to the RF signals [40]. Wavefront aberrations due to the skull had not been modeled since aberration correction techniques can be implemented efficiently at low frequencies [7,41,42].

Signal to noise ratio was calculated as the maximal power of the bubble signal located at [0, 0, 5.4cm] over the mean power of the pixels located in an area where no microbubble was located (x=8mm to 1.8mm, y=0, z=5cm to 6cm).

*2.4.2 Microvascular phantom.*

The microvasculature phantom was created using Houdini software (Side effect, Toronto) combined with Field II simulation. The phantom size was 10x10x7.5cm. The phantom was set 2.5cm away from the probe. First, a synthetic vascular network was created, three radius sizes of vessel were designed: 1.4mm, 500µm, and 200µm. Then, virtual microbubbles traveling through the vascular network were generated via Houdini software. N=100 microbubble positions were chosen randomly for each frame. Field II simulations were then computed to perform volumetric simulations. The microvasculature phantom was imaged by the three matrix arrays.

*2.4.3 Localization.*

Once the RF signals were generated and beamformed, bubble detection was performed using a standard ultrasound localization microscopy (ULM) algorithm [9]. 3D interpolation using Lanczos resampling was done to double the spatial sampling. Local maxima were detected in the image using imregionalmax Matlab function with 26 as connectivity parameter. Then for each local maxima, a Gaussian correlation was performed. Only the local maxima above -17dB (of the maximum value of the 2000 frames) were kept. The 3D Gaussian function was adapted to the size of the PSF at 5cm depth corresponding to the imaging probe. Only the positions in which the correlation coefficient was above 0.5 were considered microbubbles. Finally, a 3D paraboloid interpolation of the center of the bubble was computed.

The performance was assessed on the ability to localize the bubble with the three imaging systems. For each frame, the position assessed by the algorithm was compared to the distribution of the real bubbles set in the simulation. If the assessed position does not correspond to a real bubble to the accuracy of $\lambda/2$, the assessed position is considered as false positive. The standard deviation of microbubbles correctly localized was computed. After bubble localization, bubble positions in the 2000 frames were stacked in the imaging space to create a density map of the synthetic vascular network.

**3. Results**

*3.1 Single element*

Pressure fields from plane and curved single element transducers were estimated as a function of transducer diameter and are shown in figure 3. The $\lambda/2$ single element transducer was found quasi-omnidirectional, the pressure field transmitted by the transducer diffracts with a large angle as it can be seen qualitatively in figure 3(a) and quantitatively in figure 3(d) where its angular aperture was found to be 180°. The 4λ flat element was found to be highly directive, in figure 3(b) the pressure field is distributed near the transducer axis and an angular aperture of 50° was quantified in figure 3(d). In contrast, the curved 4λ single element transducer allowed to decrease significantly the directivity. Its angular aperture was found to be 120°.

More generally, the directivity was assessed as a function of transducer diameter for the plane and curved single element transducers. The angular aperture decreased when the size of the flat transducer increased: from 180° at $\lambda/2$ to 50° at 4λ. With curvature, angular aperture decreased but remained high: from 180° at $\lambda/2$ to 120° at 4λ figure 3(d).

The energy gain was also quantified at 10cm depth on- and off-axis as a function of diameter size for flat and curved single elements. For the flat transducer, on-axis, the gain increased in quadrature with diameter size (figure 3(e), blue curve). From $\lambda/2$ to $4\lambda$ single element diameters, the gain increased by a factor of 64. Off-axis, the gain was found to reach a maximum (gain 6) for a $2\lambda$ diameter after decreasing slightly (figure 3(f), blue curve). From $\lambda/2$ to $4\lambda$ single element diameters, the gain increased only by a factor of 5. For the curved transducer, on-axis, the gain increased with diameter size (figure 3(e), red curve). From $\lambda/2$ to $4\lambda$ diameter, the gain increased by a factor of 40. Off-axis, the gain was found to increase quasi-linearly with the diameter (figure 3(f), red curve). From $\lambda/2$ to $4\lambda$ diameter, the gain increased by a factor of 13.

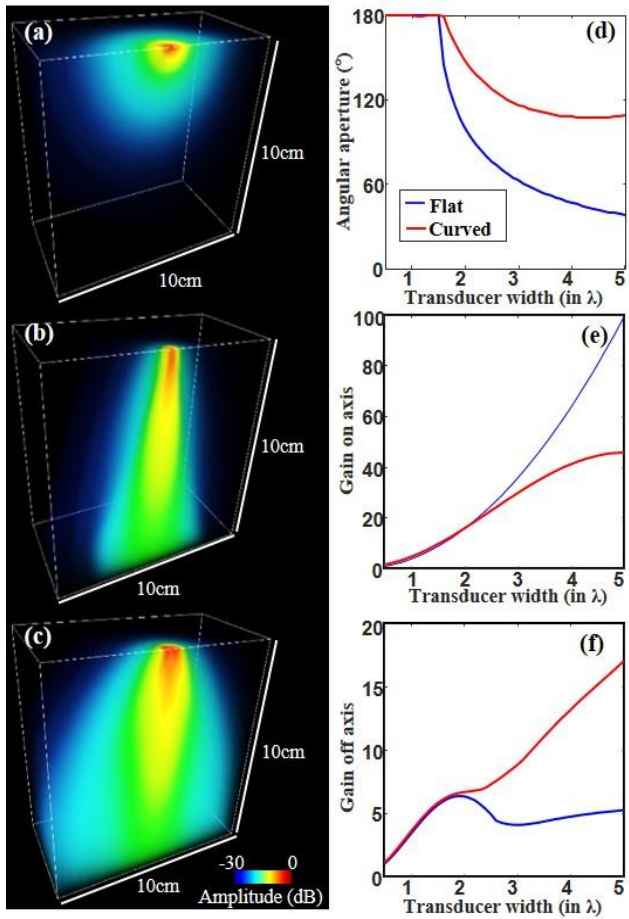

Figure 3. Comparison of flat and curved transducer emissions of different sizes. (a) Simulated pressure-field of $\lambda/2$ flat transducer. (b) Simulated pressure-field of $4\lambda$ flat transducer. (c) Simulated pressure-field of $4\lambda$ curved transducer. (d) Angular aperture as a function of transducer size (e) Gain in the imaging center as a function of transducer width. (f) Gain assessed off-axis as a function of transducer width.

On-axis, the gain is higher for plane transducers than curved transducers. However, off-axis, the gain is much lower for plane transducers than curved transducers.

### 3.2 2D array: Delay correction for curved element

A delay correction was applied to the received signal of each transducer of the 2D array to take into account the element curvature. The rephasing accuracy was assessed for the curved $4\lambda$ element transducer. RF signals backscattered from a point source at 5cm were received by a 2D array of 17 curved $4\lambda$ element transducers. RF rephased signals using classical delay law can be observed in figure 4(a). The RF signals were not found aligned meaning that the classical delay law was not appropriate for rephasing. An error defined by the maximal time shift between received signal maxima was quantified. An error of $0.65\mu s$ corresponding to 65% of the period was measured.

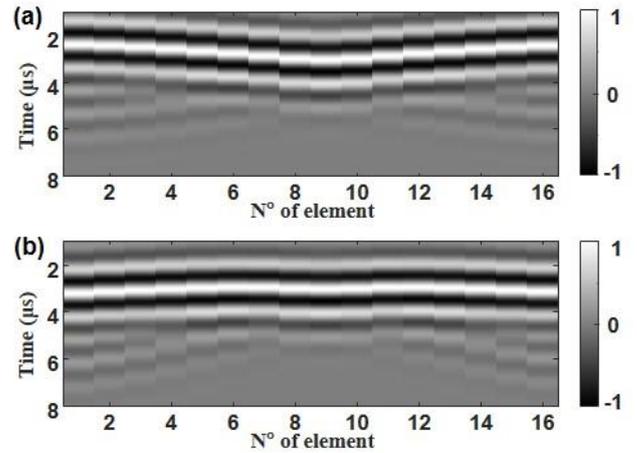

Figure 4. Uncorrected and corrected delay laws applied to the RF amplitudes, acquired by a line of curved transducers spaced of $4\lambda$. A point source had been placed at 5cm depth. (a) RF amplitudes, delayed by the focusing law computed from the apex of the curved transducer. (b) RF amplitudes, delayed by the 'corrected' focusing law computed from the focal point of the diverging lens.

Corrected delay law was used in figure 4(b) according to formula (3), RF signals were found better aligned. An error of $0.26\mu s$ corresponding to a quarter of the period was measured demonstrating a focusing enhancement.

### 3.3 Array effects: Focusing and PSF studies

#### 3.3.1 Focusing study

The probe aperture size was set to $10 \times 10$ cm² filled with 256 elements distributed with a 6mm pitch ($4\lambda$) in the two lateral axes of the probe surface. $\lambda/2$ transducers, $4\lambda$ transducers without and with curvature were compared when a focusing transmit scheme was applied. A reference matrix array composed of 16900 element transducers of half-wavelength diameter was also studied.

Focused transmit was simulated for the four configurations at the center of the field of view at a 5cm depth. The maximum pressure values were assessed.

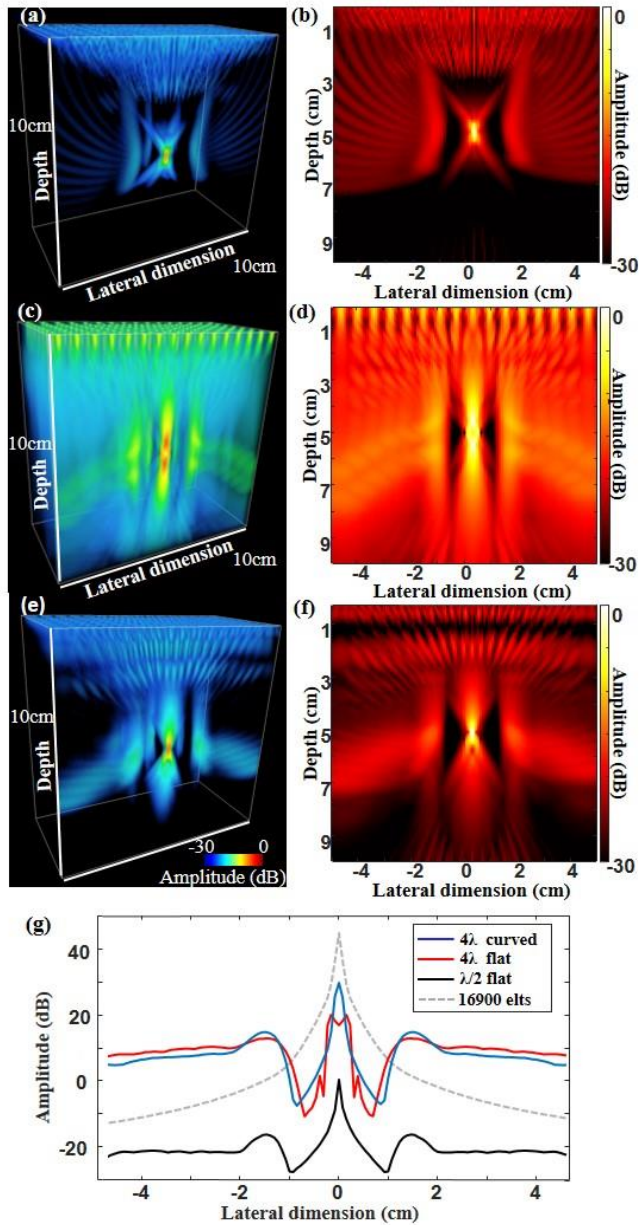

Figure 5. Focusing study comparison of matrix array probes, (a) 3D and (b) 2D maximum pressure field emitted by the λ/2 matrix array. (c) 3D and (d) 2D maximum pressure field emitted by the 4λ matrix array. (e) 3D and (f) 2D maximum pressure field emitted by the curved 4λ matrix array with the corrected focusing law. (g) Maximum pressure profiles at 5cm depth along the lateral axis.

The pressure field induced by focusing from the λ/2 matrix array was mapped and displayed in 3D (figure 5(a)) and in 2D (figure 5(b)). The lateral width of the focal spot at -6dB was found to be 1.2mm, and the level of the grating lobes was found to be -17dB with respect to its corresponding maximum peak.

The pressure field induced by focusing from the flat 4λ matrix array was mapped and displayed in 3D (figure 5(c)) and in 2D (figure 5(d)). The lateral width of the focal spot at -6dB was found to be 4.2mm, and the level of the grating lobes was found to be -7dB with respect to its corresponding maximum peak.

Finally, the pressure field induced by focusing from the curved 4λ matrix array was mapped and displayed in 3D (figure 5(e)) and in 2D (figure 5(f)). The lateral width of the focal spot at -6dB was found to be 1.8mm, and the level of the grating lobes was found to be -15dB with respect to its corresponding maximum peak.

Maximum pressure at the focus obtained with the different matrix arrays was compared in figure 5(g). The absolute pressure value at the focus using the λ/2 matrix array of 256 elements was taken as a reference (0dB). The blue, red and black lines correspond respectively to the lateral amplitude profile obtained with the 256 element array composed of 4λ curved element, 4λ flat element, and λ/2 flat element. The dashed grey line corresponds to the reference matrix array filled with 16900 elements of λ/2 flat element. Maximum pressure at the focus is 20dB, 30dB, and 45dB higher for respectively the 256 element array composed of 4λ flat element, the 256 element array composed of 4λ curved elements and the 16900 element array composed of λ/2, than the maximum pressure at the focused for the 256 element array composed of λ/2 elements.

### 3.3.2 Point spread function

The point spread functions (PSF) at the location [0, 0, 5cm] were compared using the matrix arrays defined above. To perform ultrafast imaging, nine transmits were emitted successively. Coherent compounding, and classical 3D delay and sum algorithm were applied to the received signal to beamform a PSF volume. The PSF corresponding to the matrix array using λ/2, 4λ flat and 4λ curved elements are displayed in 3D respectively in figure 6(a), (c) and (e), the corresponding maximum projection on the plane Oxy are displayed respectively in figure 6(b), (d), and (f), and finally, the corresponding maximum projection on the lateral axis are displayed respectively in figure 6(g), black line, red line, and blue line.

The focal spot of the PSF corresponding to the λ/2 matrix array and the 4λ curved matrix array were found well-localized contrary to the focal spot of the PSF corresponding to the 4λ matrix array. The lateral width of the focal spot at -6dB was assessed: 1.2mm, 4.2mm and 1.8mm was measured respectively for the PSF corresponding to the λ/2, 4λ flat and 4λ curved matrix array, as well as the secondary lobe level respectively at -28dB, -7 and -17dB.

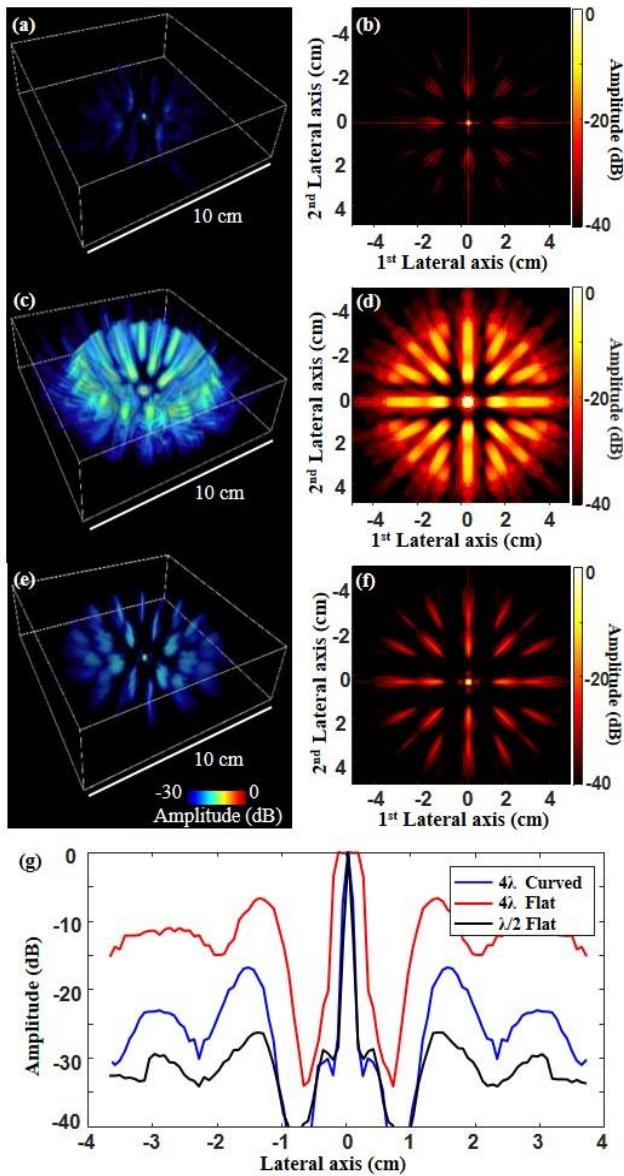

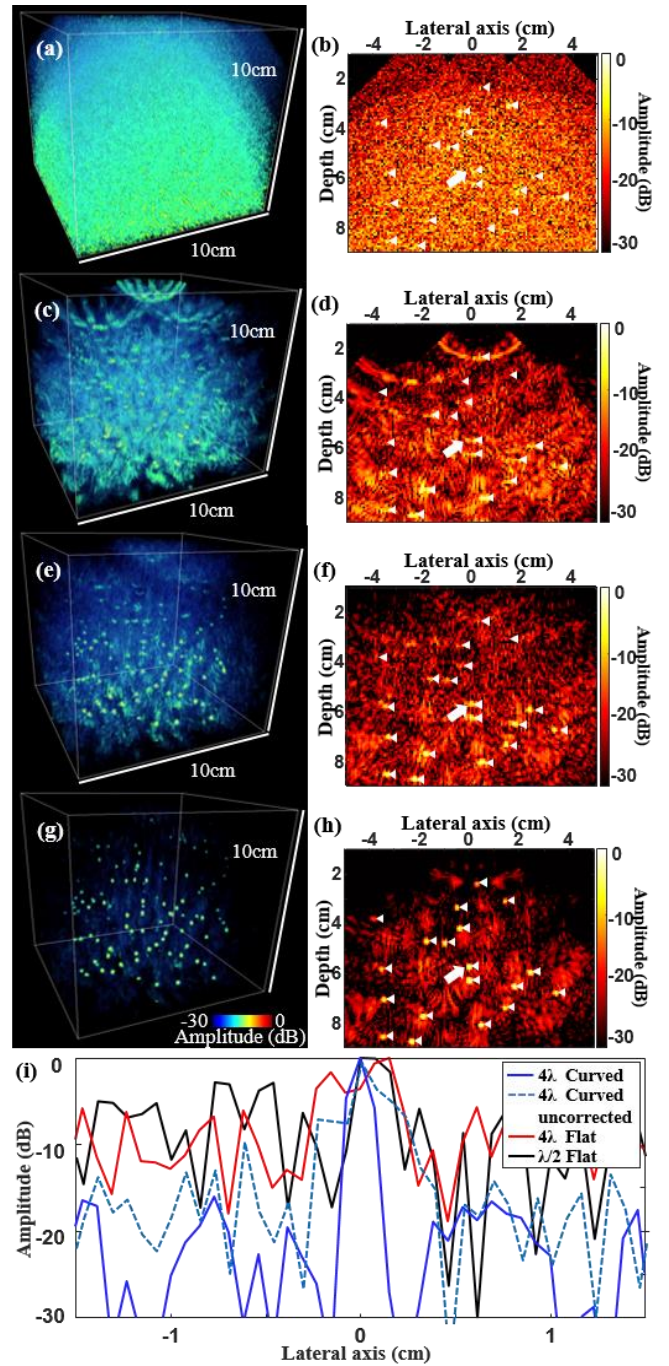

Figure 6. Point spread function (PSF) of matrix array probes. (a) 3D PSF obtained with the probe composed of λ/2 elements (b) and its projection. (c) 3D PSF obtained with the probe composed of 4λ flat elements (d) and its maximum projection. (e) 3D PSF obtained with the probe composed of 4λ curved elements (f) and its maximum projection. (g) Maximal values projected on the x-axis.

## 3.4 Imaging and ultrasound localization microscopy performances uncorrected

### 3.4.1 Static phantom imaging

3D B-mode volumetric imaging of the static phantom obtained with the λ/2 diameter element matrix array is displayed in figure 7(a). Bubbles are barely distinguishable due to a low signal-to-noise ratio that worsens with depth. Oxz plane is also displayed in figure 7(b). Few spots corresponding to bubbles are observed between the depth of 1cm and 4cm. The black line in figure 7(i) displays the intensity profile of a bubble located at [0, 0, 5.4cm] pointed by a white arrow in figure 7(b). The image in figure 7(b) is very noisy near the location of the white arrow and the microbubble is barely distinguishable, as well as the corresponding intensity profile (two pics corresponding to noise signals can be seen only at -3dB of the maximum amplitude corresponding to the microbubble signal). A maximum can be seen at x = 0cm which corresponds to the location of the bubble. The signal-to-noise ratio (SNR) was measured and 10dB was found.

Figure 7. Volume of 200 scatters randomly distributed in a 10x10x10cm volume using 2D matrix probes composed of 256 elements. 3D image obtained using a probe composed of λ/2 transducers (a), 4λ flat transducers (c), 4λ curved transducers with uncorrected delay laws (e), 4λ curved transducers with corrected delay laws (g), the respective Oxz plane (b), (d), (f), (h). (i) The x-profile amplitude of the bubble, located at [0,0,5.4cm], and indicated by the white arrow in figure (b) black line, (d) red line, (f) light dashed blue and (h) blue line. The left corner of white triangles on (b), (d), (f), and (h) points out the positions of the microbubbles in the Oxz plane.

3D B-mode volumetric imaging obtained with the 4λ diameter element matrix array is displayed in figure 7(c). Bubbles are distinguishable deep in the field of view but with a low SNR and inaccurate location detection, distorting the corresponding vessels. The Oxz plane is shown in figure 7(d) and confirmed previous observations. The red line in figure 7(i) displays the intensity profile of a bubble located at [0, 0, 5.4cm] pointed by a white arrow in figure 7(d). A maximum can be seen at x = 0.2cm which corresponds to the location of the bubble. The width of the maximum at -6dB was measured (5mm) as well as the SNR (12dB).

3D B-mode volume obtained with the curved 4λ diameter element matrix array with uncorrected delay laws is displayed in figure 7(e). Several intensity spots corresponding to bubbles are distinguishable. The Oxz plane is shown in figure 7(f). The light dashed blue line in figure 7(i) displays the amplitude profile of a bubble located at [0, 0, 5.4cm] pointed by a white arrow in figure 7(f). A maximum can be seen at x = 0cm which corresponds to the location of the bubble. The width of the maximum pic at -6dB was assessed: 2.9mm. The signal-to-noise ratio (SNR) was measured: 17dB.

3D B-mode volume obtained with the curved 4λ diameter element matrix array with corrected delay laws is displayed in figure 7(g). Clear intensity spots corresponding to bubbles are distinguishable. The Oxz plane is shown in figure 7(h) and confirmed previous observations. The blue line in figure 7(i) displays the intensity profile of a bubble located at [0, 0, 5cm] pointed by a white arrow in figure 7(h). A maximum can be seen at x = 0cm which corresponds to the location of the bubble. The width of the maximum pic at -6dB was assessed: 2mm. The signal-to-noise ratio (SNR) was measured: 25dB.

### 3.4.2 Microcirculation phantom imaged with ultrasound localization microscopy

A transcranial vasculature dynamic phantom with 200 microbubbles randomly distributed was emulated with Houdini software and imaged with the previously studied matrices using Field II software. A 3D ULM algorithm was used to localize the bubble position in each volume as explained in the method section 2.4.2. By cumulating the bubble position over 2000 volumes, a density map of the vascular phantom was constructed. The ground truth density map of the phantom is displayed in figure 8(a). Density maps were constructed for different 256 element matrix arrays: the λ/2 element matrix array in figure 8(b), the 4λ flat element matrix array in figure 8(c), the 4λ curved element matrix array without delay correction in figure 8(d), and the curved 4λ element matrix array with delay correction in figure 8(e).

Bubble detection performances were assessed for the different matrix arrays and are gathered in table 2. For the λ/2 element matrix array, only 7% of the microbubbles were correctly localized. The detected bubbles were mostly located near the probe between 2.5cm and 4cm depth due to a poor SNR. For the 4λ element matrix array, 10% of the microbubbles were accurately detected. The bubbles were mostly located below 6 cm depth. Near the probe, vessel network was missing because the bubbles were not detected. Large flat elements are very directive and angles are too wide to enable good focusing quality which is necessary to image the bubbles. This effect can be observed in the first 6 cm in figure 7(d). A large number of detected bubbles were found to be false positive (70%). For the curved 4λ element matrix array without delay correction, the density map reveals the phantom vasculature but only after 5 cm depth. 4% of the microbubbles were accurately detected. Near the probe, the vessel network was missing because the bubbles were not detected. When delay laws are not corrected, errors in the focusing delay laws occur, which leads to degraded microbubble images. These errors depend on depth and are more important near the probe. This effect can be assessed in figure 7(f). A large number of detected bubbles were found to be false positive (90%) because uncorrected delay laws induce non-uniform shifts in the image, thus microbubbles are detected at wrong locations.

The 3D density map obtained using the matrix with 4λ curved elements combined with corrected delay laws is displayed in figure 8(e). The density map revealed the whole phantom with vessels from 1.4mm down to 200µm. Figure 8(e) and figure 8(a) are very similar. 93% of the microbubbles were accurately detected, with very few false positives (<1%).

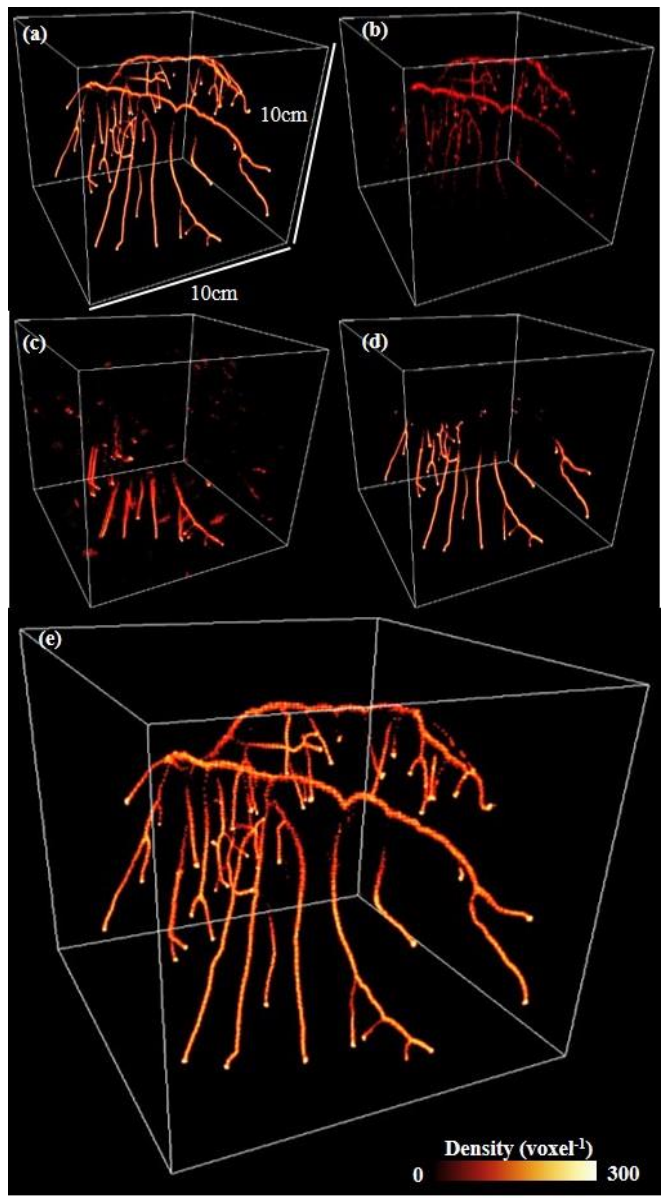

Figure 8. 3D Ultrasound localization microscopy (ULM) density map of 2000 frames of a vascular phantom using different 2D matrix array probes composed of 256 elements. (a) Ground truth. ULM density map obtained with λ/2 element (b), and 4λ flat element (c) matrix array. ULM density map obtained with 4λ curved element matrix array without corrected delay laws (d) and with corrected delay laws (e).

TABLE 2

| Probe used | Bubble correctly detected | Std | False positive |
|---|---|---|---|
| Flat λ/2 | 7 % | 2.0% | 3% |
| Flat 4λ | 10 % | 2.2% | 74% |
| Curved 4λ, uncorrected delay laws | 4% | 1.6% | 90% |
| Curved 4λ, corrected delay laws | 93% | 2.7% | 0% |

Table 2: Dynamic phantom quantification.

## 4. Discussion

In this study, a new approach for transcranial 3D ULM based on the concept of an ultrasensitive multi-lens probe with a reduced number of large elements was proposed and studied via simulations. The performances of the multi-lens probe with large elements had been made possible by using delay correction laws, mandatory for optimal focusing. The 256 element probe, driven at low acoustic frequency was able to perform 3D ULM of a complex vasculature network through the skull bone in a large field of view.

The influence of a negative acoustic lens modeled as a curved transducer was first studied on a single element as a function of transducer diameter. The results on a single element showed that large curved transducers of 4λ in diameter were able to maintain a large angular aperture (120°) while increasing the pressure by 40 times on-axis and 13 times 30° off-axis, compared to lambda/2 transducers. The imaging performances were assessed for the different transducer arrays for ultrafast imaging. We demonstrated that the array composed of 256 4λ curved elements was able to provide at the same time high pressure amplitude, low grating lobes, and good resolution when compared to the other arrays composed of λ/2 transducer and 4λ flat transducers.

The capability of the three sparse matrix array probes to image unique scatterers in a volume was tested with the presence of noise and pressure amplitude attenuation induced by the skull bone. Among all tested configurations, the only configuration which could successfully detect the scatterers with a high SNR, in the whole phantom volume, was found for the probe composed of 4λ diameter transducers with curvature. The feasibility of performing 3D ULM through the skull bone on a realistic flow particle phantom of a complex vasculature network with micro-bubble injection was assessed for the different configurations. Again, the probe composed of 4λ diameter transducers with curvature was able to successfully detect the bubbles with excellent accuracy (>90%) and reconstruct in 3D the vasculature network using 3D ULM techniques.

The multi-lens probe with large elements, modeled as curved transducers in this study, presents many advantages. It enabled transmitting and receiving high acoustic amplitude due to its large element area. It enabled maintaining a low directivity which increased image/volume quality in terms of field of view, resolution, and sensitivity. The multi-lens probe with large elements was introduced in the field of photoacoustics [34] for those advantages. Only used in receive mode, we extended this concept to transmit/receive ultrasound imaging.

The acoustic energy loss due to reflection, multiple scattering, and absorption of ultrasound waves in cranial bone reduces highly the sensitivity and currently limits imaging through very particular windows such as the temporal bones to image the brain or between the ribs to image the heart or the liver. The proposed configuration promises to alleviate current challenges in the ultrasound imaging field when imaging behind bones. These windows of few cm² limit aperture size and thus resolution and sensitivity. The use of acoustic lenses or curved transducers could be key to enlarge aperture size to image organs in large volumes through the bones while keeping a high sensitivity. Wavefront aberrations can be expected due to the propagation of the ultrasonic wave through the bones. However, at low frequencies aberration is limited and can be modeled as a near-field phase screen. Additional time delays can then be implemented on each element of the matrix array to overcome this issue [41,42], as it was done in the clinical study made by C. Demene, J. Robin et al [7].

Furthermore, the use of large elements provides a large active transducer surface while reducing significantly the number of elements. In this study, only 256 elements were used for an aperture size of 10x10cm enabling imaging a volume of 10x10x10 cm which could correspond to image the whole human brain. The equivalent array fully populated with small elements would require 16900 piezoelectric elements implying heavy technology and tremendous cost which would limit its clinical translation.

The use of large elements imposes a large pitch between the elements of the matrix array and induces important grating lobes. High grating lobe level was measured (-17dB) with the 4λ curved matrix array. However, despite this high level of grating lobes, we were able to perform microbubble localization with excellent performances (93% microbubbles detected) for ultrasound localization microscopy. Decreasing the pitch from 4λ to 3λ would reduce the grating lobe level but at the cost of a larger number of elements (two times more). Further decreasing the spatial pitch down to 2λ would increase the number of elements by a factor of 4 while reducing the interest of the multi-lens approach compared to flat elements.

The multi-lens diffracting layer enables acquiring a large field of view in three dimensions. Like other 3D ultrasound imaging methods, large and small motion can be assessed in three dimensions. It solves out-of-plane motion issue which can occur when using two-dimensional probe arrays. Different strategies such as phase cross-correlation [43] or 3D speckle tracking [44] could be envisioned to solve complex multi-scale motions that can occur during images acquisition.

Interesting ULM studies [45,46], had been carried out to map the brain vascular network with a reduced number of transducers by using a sparse array made of small elements (<1λ of diameter). However, to overcome sensitivity issues due to small element size, the sparse elements are located on a spherical surface, which limits the reachable field of view to a small volume (typically $12 \times 12 \times 12$ mm) surrounding the geometrical focal spot [46].

To our knowledge, there is no existing solution able to image the whole brain at a high volume rate using a limited number of transducers and able to overcome sensitivity problems associated with transcranial imaging. Our imaging system had been conceived to fulfill those requirements, and we proved via simulation its imaging capabilities.

Transcranial multiscale and functional evaluations of microvasculature at the bedside of the patients would be a major improvement of patient care, as it addresses a major unmet clinical need. The portability, low cost, and non-ionizing features of ultrasound technology for cerebral angiography and hemodynamics mapping at much higher resolutions than the current state-of-the-art MR and CT Angiography could become a strong added value for clinical translation. Such microangiographic imaging could be envisioned in the diagnostic of stroke and aneurysms as well as the monitoring of post-stroke recovery and aneurysms survey. For cancer, the 3D high-resolution mapping of brain tumors could help to better define resection margins and improve surgical planning. It could also offer new insights into the fundamental understanding and characterization of complex lesions such as glioblastoma multiform. It could also be applied to the microcirculation of any organs for human patients or animal models and will be particularly interesting for deep organs with rapid motion including the heart, the liver, or the kidney.

The correction delay law described in this study is essential to the use of multi-lens probes. It relies on the principle that the backward wave travels toward the effective curvature radius of the lens and not toward the element center as it is done in classical beamforming. This correction, performed in transmit/receive signals, enabled to restore a well-aligned focusing law (figure 4(b)). The study on the dynamic phantom demonstrated the extreme importance of this correction law as the vasculature network could not be reconstructed without it.

The dynamic phantom modeled with Houdini software promises to provide important information for ULM applications. Parameters such as the number of particles, tube length, tube diameter, particle velocity, and hydrodynamic laws were fully tunable. The software gave access to particle positions as well as instant velocities at each frame defined by the user. These data were loaded in Field II and imaged with our probes. The coordinates measured after localization through a standard 3D ULM algorithm were compared to the true coordinates given by Houdini software. This framework could be extended to any probe configurations to demonstrate ULM feasibility. Furthermore, it could be extremely useful to assess ULM algorithm performances in terms of localization and tracking by comparing measured velocity with ground truth. Ongoing work aims to further explore the capability of this proposed framework.

In Field II simulation software, transducers with lenses were modeled by curved transducers which is a limitation. Complex reflection at the interface between the lens material and the piezoelectric component could occur and increase the noise level. Addition noise can also occur because of cross-talk between nearby transducers and parasite side reflections. However, this type of noise being deterministic deconvolution methods [47] could be used to remove it.

In transmit, plane or diverging wave transmits could not be used due to low element transducer spatial sampling creating inhomogeneity in the pressure field. Ultrafast imaging was performed by emitting ultrasonic waves by one element at a time similarly as it is done in synthetic aperture imaging techniques [48]. It limited the transmit energy because only one element was used instead of the entire aperture. It was partially compensated by the size of a single element in the proposed method. However, if needed coded emissions with multiple transducers per transmit could be implemented to increase SNR [49,50].

The time needed to process the data depends on the imaging field, the number of microbubbles in the simulation, and the computation power. In our setup, the simulation code was not parallelized and was performed, using Matlab, on a CPU (AMD Ryzen5 3600X 6-core processor, 3.79 GHz). It took 5s to generate the Radio Frequency signals of 100 microbubbles and 27s to beamform one volume of 2.1 million voxels (corresponding to a $10 cm^3$ volume sampled at $(\lambda/2)^3$). The localization of the microbubbles over 2000 frames took 3min 45s. GPU parallelization could be implemented to reduce the beamforming time as it is done in [19].

Ultrafast imaging is an important asset for our application in vivo: for microbubble tracking to assess velocities, but also to enhance microbubble SNR in post-processing. As moving microbubbles and tissue backscatter signals have different spatial coherence signatures, the tissue signals can be removed using singular decomposition value filtering (SVD) [51], and thus SNR of Microbubble signals can be enhanced. The SVD filtering required hundreds of frames to work efficiently and a high volume rate is mandatory.

In vivo applications are more complex and safety measurements will be necessary for human applications. These studies go beyond the scope of this manuscript and will be addressed in future work.

## 5. Conclusion

We proved the imaging capabilities, for transcranial ultrafast localization microscopy application, of a new ultrasensitive probe with 256 large diverging elements that can be performed using a multi-lens diffracting layer or curved transducers. The idea of this ultrasensitive probe composed of a limited number of transducers is to boost the sensitivity by increasing the surface of the single transducer and to overcome the following increase of directivity by adding a diverging element. A corrected delay law, essential to form images had successfully been applied to image a 3D dynamic phantom.

The proposed multi-lens diffracting layer has a strong potential to enable 3D ULM over a large field of view through the bones.

## Acknowledgements

This work was supported by the AXA research found and Inserm research accelerator (Inserm ART).